# MultiRefactor: Automated Refactoring To Improve Software Quality

Michael Mohan (0000-0002-9944-5821)✉ and Des Greer

Queen's University Belfast, Northern Ireland, UK
`{mmohan03, des.greer}@qub.ab.uk`

**Abstract.** In this paper, a new approach is proposed for automated software maintenance. The tool is able to perform 26 different refactorings. It also contains a large selection of metrics to measure the impact of the refactorings on the software and six different search based optimization algorithms to improve the software. This tool contains both mono-objective and multi-objective search techniques for software improvement and is fully automated. The paper describes the various capabilities of the tool, the unique aspects of it, and also presents some research results from experimentation. The individual metrics are tested across five different codebases to deduce the most effective metrics for general quality improvement. It is found that the metrics that relate to more specific elements of the code are more useful for driving change in the search. The mono-objective genetic algorithm is also tested against the multi-objective algorithm to see how comparable the results gained are with three separate objectives. When comparing the best solutions of each individual objective the multi-objective approach generates suitable improvements in quality in less time, allowing for rapid maintenance cycles.

## 1 Introduction

Search based optimization has been used extensively in various areas of engineering and in recent years has also been applied to software engineering. Search Based Software Engineering (SBSE) is an area of research that attempts to apply search heuristics to solve complex problems in software development [1]. Software maintenance is one of the more expensive parts of the software development cycle [2]. SBSE applied to maintenance, known as Search Based Software Maintenance (SBSM), is used to assist the manual aspects of maintaining a software project and minimize the time necessary to do so. To aid with this research various tools [3–11] have been used to assist with the refactoring of a software project. An increasing amount of SBSM research is looking at multi-objective techniques [12–20]. Many multi-objective search algorithms are built with genetic algorithms, as their ability to generate multiple possible solutions is suitable for a multi-objective approach. Instead of focusing on only one property, the multi-objective algorithm will be concerned with a number of different objectives.

The MultiRefactor tool uses refactorings to improve Java projects using metric functions to guide the search. Many of the other tools available have a limited selection of refactorings or metrics available to use. The effort has been made to equip the MultiRefactor tool with a large range of available refactorings and metrics to choose from, in order to promote maximum configurability within the tool. MultiRefactor combines the ability to use a multi-objective approach with the more practical ability to improve the source code itself, while checking the semantics of the refactorings being applied so that the changes in the code are valid with respect to the application domain.

In order to assess the capabilities of the MultiRefactor approach, a set of experiments have been set up to compare different procedures available within the tool. Experiments have previously been conducted comparing the other metaheuristic searches [21], so the experimentation here focuses on the use of the genetic algorithms in the tool and aims to find out two things. The first aim is to test the available software metrics within the tool and discover which are more successful. Some metrics may be more useful than others in measuring the changes made by the available refactorings. These will be more helpful when trying to analyze the changes made to a solution and as such, a metric function made from these metrics may assist in creating a more prosperous solution. The second aim is to compare the mono-objective approach with the multi-objective search available and see whether using a multi-objective algorithm to automate maintenance of a software solution is as practical as using a mono-objective algorithm. We wish to test whether, in a fully automated solution, a multi-objective algorithm using similar settings can yield comparable results across all the objectives, and whether it is worth the time taken to do so. The following research questions have been formed to address these concerns, along with a corresponding set of hypotheses and null hypotheses for each factor investigated in **RQ2**:

**RQ1:** Which set of software metrics have the most variability when used with a mono-objective genetic algorithm to refactor software?

**RQ2:** Does a multi-objective refactoring approach give comparable results on all objectives to corresponding mono-objective refactoring runs?

**H1:** The overall objective improvements in the multi-objective searches are not significantly worse than the overall objective improvements in the mono-objective search.

**H1$_0$:** The overall objective improvements in the multi-objective search are significantly worse than the overall objective improvements in the mono-objective searches.

**H2:** The overall time taken to run the multi-objective search is no higher than the time taken to run any of the three mono-objective searches.

**H2$_0$:** The overall time taken to run the multi-objective searches is higher than time taken to run one of more of the three mono-objective searches.

The remaining sections go into more detail about the capabilities of the MultiRefactor approach and showcase its abilities with the set of experimental studies. Section 2 discusses the design of the tool as well as the refactorings, metrics and search techniques available. Section 3 explains the details of the experiments conducted. The results are presented in Section 4 and discussed in Section 5. Section 6 presents related literature within SBSE and with multi-objective techniques in SBSM. Finally, Section 7 gives the conclusion.

## 2 MultiRefactor

The MultiRefactor approach[1] is in common with those of Moghadam and O' Cinnéide [10] and Trifu et al. [7] in using the RECODER framework[2] to modify source code in Java programs. RECODER extracts a model of the code that can be used to analyze and modify the code before the changes are applied and written to file. The tool takes Java source code as input and will output the modified source code to a specified folder. The input must be fully compilable and must be accompanied by any necessary library files as compressed jar files. The numerous searches available in the tool have various input configurations that can affect the execution of the search. The refactorings and metrics used can also be specified. As such, the tool can be configured in a number of different ways to specify the particular task that you want to run. If desired, multiple tasks can be set to run one after the other.

A previous study [22] used the A-CMA [9] tool to experiment with different metric functions but needed to be modified to produce an output. The tool could only produce bytecode (likewise, the TrueRefactor [3] tool only modifies UML and Ouni et al.'s [17] approach only generates proposed lists of refactorings) so the MultiRefactor tool was developed in order to be a fully automated search-based refactoring tool that produces compilable, usable code as an output. The tool can therefore be used for research purposes or for maintaining actual projects, as demonstrated in Section 3 where open source projects are used for experimentation. Along with the Java code artifacts, the tool will produce an output file that gives information on the execution of the task. The output gives information about the parameters of the search executed, the metric values at the beginning and end of the search, and details about each refactoring applied. The metric configurations can be modified to include different weights and the direction of improvement of the metrics can be changed depending on the desired outcome. These configurations can be read in a number of ways including as text files or xml files. There are a few ways the metrics functions can be calculated. An overall metric value can be found using a weighted metric sum or Pareto dominance can be used to compare individual metrics within the functions. Fig. 1 gives a brief overview of the process used in the MultiRefactor tool to generate refactored Java code.

---

[1] https://github.com/mmohan01/MultiRefactor
[2] http://sourceforge.net/projects/recoder

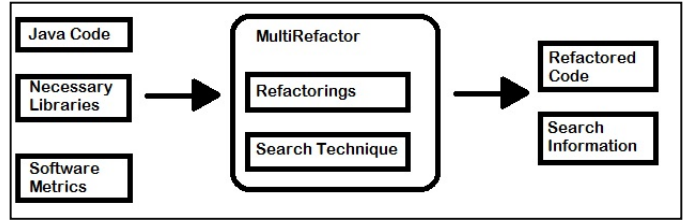

**Fig. 1.** Overview of the MultiRefactor process

**2.1 Searches**

MultiRefactor contains six different search options for automated maintenance, with three distinct metaheuristic search techniques available. For each search type there is a selection of configurable properties to signify how the search will run. For the searches used in this paper (the genetic algorithm and the multi-objective genetic algorithm) the details of how they are implemented and the configurable properties available are given below.

The *Genetic Algorithm* is based on the process of genetic replication. The representation used in MultiRefactor is based on the implementation used by Seng et al. [23] and further adapted by O' Keeffe and O' Cinnéide [24]. The search algorithm stores model information to represent multiple different genomes in a population, avoiding the expensive memory costs needed to store multiple different models. The initial population is constructed by applying a selection of random refactorings to the initial model to create a single genome, and repeating for the required amount. The crossover process uses the cut and splice technique, generating two offspring from two different parent genomes. A single, separate point is chosen for each parent in order to facilitate the technique. The point is chosen at random along the refactoring sequence in each of the parent solutions, with at least one refactoring present on each side. For each child, the two sets of refactorings are then mixed together. The first set of refactorings in one parent will be applied first and then the second set of refactorings from the other parent will be applied. Any inapplicable refactorings during this process will be left out although the child genome will still be able to be generated using the remaining refactorings. Mutation will choose from the new offspring and apply a single random refactoring to the end of the refactoring sequence for that genome. Crossover will be applied at least once during each generation and may happen more depending on the input parameters specified. Likewise, mutation will be applied a certain amount of times each generation depending on the parameters specified, or may not happen at all.

In order to choose parent genomes for crossover, a rank selection operator is used. Once the mutation process is complete for a generation, the new offspring is combined with the current population and the solutions are ordered according to fitness. The genetic algorithm can either store the entire final population of solutions resulting from the process, or only the fittest solution. The amount of generations specified will determine when the search terminates and the population size will determine how many genomes are generated during initialization and how many will survive each generation. The crossover probability and mutation probability (between 0 and 1) determine the likeliness of these processes being executed during the search. The refactoring range will determine the initial amount of refactorings ap-plied to the genomes during the initialization process. For each initial solution, a random amount of refactorings between one and the refactoring range will be chosen.

The *Multi-Objective Genetic Algorithm* is largely identical to the simple genetic algorithm, and contains the same configuration options (although it must store the whole population when finished). The algorithm is an adaptation of the NSGA-II [25] algorithm and as such, differs mostly in how the fitness is calculated. The selection operator used is the binary tournament operator, in order to avoid the need to rely on ranks during selection. The multi-objective algorithm allows the user to choose multiple metric functions as separate objectives to guide the search. The genomes in the population will then be sorted using a non-dominated approach, allowing each objective to be considered separately. Unlike the approach used by Ouni et al. [17], the refactorings used will be checked for semantic coherence as part of the search, and will be applied automatically, eliminating the need to check and apply the refactorings manually and ensuring the process is fully automated. There is also a many-objective search available in the tool to handle problems with more than three objectives.

**2.2 Refactorings**

The refactorings used in the tool are mostly based on Fowler's list of refactorings [26], and consist of 26 field-level, method-level and class-level refactorings, as listed in Table 1. Each refactoring will initially deduce whether a program element can be refactored. It will make all the relevant semantic checks and return true or false to reflect whether it is applicable as a refactoring and whether the code will be able to compile after it is applied. The checks applied will depend on the refactoring, and are important in order to exclude elements that are not applicable for that refactoring. These checks, as well as the refactoring process itself, ensure that the refactorings chosen are behavior preserving, and that the program will still be compilable after the refactorings are applied to the solution. The RECODER framework allows the tool to apply the changes to the element in the model. This may consist of a single change or, as in the case of the more complex refactorings, may include a number of individual changes to the model. Specific changes applied with the RECODER framework consist of either adding an element to a parent element, removing an element from a parent element, or replacing one element with another in the model. The refactoring itself will be constructed using these specific model changes.

**Table 1.** Available refactorings in MultiRefactor tool

| Field Level | Method Level | Class Level |
| --- | --- | --- |
| Increase Field Visibility | Increase Method Visibility | Make Class Final |
| Decrease Field Visibility | Decrease Method Visibility | Make Class Non Final |
| Make Field Final | Make Method Final | Make Class Abstract |
| Make Field Non Final | Make Method Non Final | Make Class Concrete |
| Make Field Static | Make Method Static | Extract Subclass |
| Make Field Non Static | Make Method Non Static | Collapse Hierarchy |
| Move Field Down | Move Method Down | Remove Class |
| Move Field Up | Move Method Up | Remove Interface |
| Remove Field | Remove Method | |

In some cases new elements will be created for use in the refactoring (for instance, new imports may need to be created when moving an element to a new class), and where possible, these will be constructed from existing elements to minimize the potential for issues. The refactorings can be reversed to undo the changes made in the last instance of the refactoring. This allows the hill climbing and simulated annealing searches to check neighboring refactorings from the current state and measure their impact on the program, before deciding which one to use. For some refactorings, choices have to be made in relation to how specifically the refactoring is applied. The Move Field Down and Move Method Down refactorings involve moving program elements down to a sub class. Here, the subclass to be used needs to be chosen before the refactoring is applied. Likewise, the Extract Subclass refactoring involves picking a subset of the elements of a class to extract into a new sub class. Here the elements to be moved will need to be chosen beforehand.

The *Increase/Decrease Visibility* refactorings change a field or method declaration up or down one level between public, protected, package and private visibility (where an increase moves towards private and a decrease moves towards public). The *Make Final/Non Final* refactorings will either apply or remove the final keyword from a field, method or class declaration. Likewise, the *Make Static/Non Static* refactorings are concerned with added or removing the static keyword from a global field or method declaration. Also, *Make Class Abstract/Concrete* will add or remove the abstract keyword from a class declaration. The *Move Down/Up* refactorings will either move the global field or method declaration to its immediate super class or to one of its available sub classes. *Extract Subclass* will choose a selection of local field and/or method declarations from a class that relate to each other as a distinct unit, and will move them to a newly created sub class. *Collapse Hierarchy* is applied by taking all the elements of a class (except any existing constructors for the class) and moving them up into the super class. It will then remove the class from the hierarchy. The *Remove* refactorings will remove the element related to that type of refactoring.

**2.3 Metrics**

The metrics in the tool are used to measure the current state of a program and deduce whether an applied refactoring has had a positive or negative impact. Due to the multi-objective capabilities of MultiRefactor, the metrics can be measured as separate objectives to be more precise in measuring their effect on a program. A number of the metrics available in the tool are adapted from the list of metrics in the QMOOD [27] and CK/MOOSE [28] metrics suites. Table 2 lists the 23 metrics currently available in the tool and the metrics not adapted from elsewhere are described below.

**Table 2.** Available metrics in MultiRefactor tool

| QMOOD Based Metrics | CK Based Metrics | Others |
|---|---|---|
| Class Design Size | Weighted Methods Per Class | Abstractness |
| Number Of Hierarchies | Number Of Children | Abstract Ratio |
| Average Number Of Ancestors | | Static Ratio |
| Data Access Metric | | Final Ratio |
| Direct Class Coupling | | Constant Ratio |
| Cohesion Among Methods | | Inner Class Ratio |
| Aggregation | | Referenced Methods Ratio |
| Functional Abstraction | | Visibility Ratio |
| Number Of Polymorphic Methods | | Lines Of Code |
| Class Interface Size | | Number Of Files |
| Number Of Methods | | |

*Abstractness* measures the ratio of interfaces in a project over the overall amount of classes. *Abstract Ratio* gives the average ratio of abstract methods (as well as the class itself if it is abstract) per class. *Static Ratio* and *Final Ratio* give the average ratios of static and final elements per class (static amount looks at classes and methods, whereas final amount also looks at fields), and *Constant Ratio* calculates the average ratio of elements (classes, methods and global fields) that are both static and final pre class. *Inner Class Ratio* calculates the ratio of the amount of inner classes over the amount of classes in a project. *Referenced Methods Ratio* finds the average ratio of inherited methods referenced per class. In each class, the metric measures the amount of distinct external methods (methods defined outside the current class) referenced amongst the methods of the class. For each class, the ratio of the amount of these methods that are inherited by the class over the amount referenced is calculated. *Visibility Ratio* calculates an average visibility ratio per class. In a class, each method and global field declaration (as well as the class itself) is given a visibility value, where a private member has a value of 0 and a public member has a value of 1. The visibility ratio for that class will calculate the accumulated visibility values over the amount of elements. The smaller this is, the more inaccessible the elements of the project are. Finally, *Lines Of Code* gets the overall amount of lines of code in a project and *Number Of Files* counts the amount of Java files in a project.

# 3 Experimentation

Five open source programs are used in the experimentation to ensure a variety of different domains are tested. The programs range in size from relatively small to medium sized, as shown in Table 3. These programs were chosen as they have all been used in previous SBSM studies and so there is an increased ability to compare the results and also because they promote different software structures and sizes. The source code and necessary libraries for all of the programs are available to download in the GitHub repository for the MultiRefactor tool. The experiments are run on a PC using an Intel Core i7 CPU and with 8GB of RAM. The experimentation is split into two parts. The first experiment measures the effect of each individual metric available on a range of inputs using the mono-objective genetic algorithm. The second experiment compares the more effective metrics in a mono-objective set up against a multi-objective approach. In order to choose configuration parameters for the genetic algorithms used, trial and error is used to find the most effective settings. First, the crossover and mutation probabilities are compared using a baseline metric and input. The largest input, JHotDraw, is used with a metric assumed to be volatile due to it being directly related to the increase/decrease visibility refactorings, visibility ratio. Nine different tasks are used to compare crossover and mutation probabilities of 0.3, 0.5 and 0.8. Each task is run five times to get an average value. As shown in Fig. 2, the most improved configuration has a mutation value of 0.8 and a crossover value of 0.2.

**Table 3.** Java programs used in experimentation

| Name | LOC | Classes |
|---|---|---|
| JSON 1.1 | 2,196 | 12 |
| Mango | 3,470 | 78 |
| Beaver 0.9.11 | 6,493 | 70 |
| Apache XML-RPC 2.0 | 11.616 | 79 |
| JHotDraw 5.3 | 27,824 | 241 |

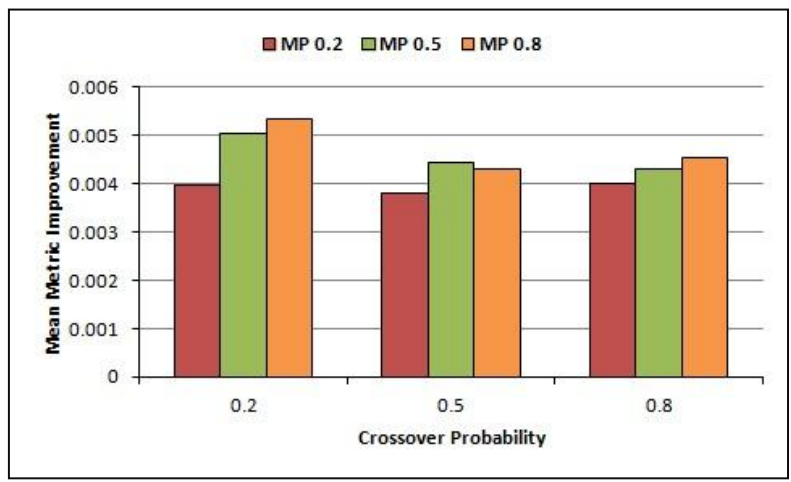

**Fig. 2.** Mean metric improvement values with different crossover and mutation probabilities

Next, the other configuration parameters are compared using these mutation and crossover values to find the best tradeoff between software improvement and time taken. 27 different tasks are set up to compare different combinations of generation amounts, population sizes and refactoring ranges. The generation amounts tested are 50, 100 and 200. The refactoring ranges used are likewise and the population sizes used are 10, 50 and 100. Fig. 3 shows the metric improvement values for each permutation of the generation, population size and refactoring range genetic algorithm settings. Fig. 4 compares them against the time taken to run them. As shown in Fig. 4, one configuration stands out as having a larger increase in quality without having a similar increase in necessary time. This configuration with 100 generations, a refactoring range of 50 and a population size of 50 is used for the experimentation. The final settings are shown in Table 4.

**Table 4.** Genetic algorithm configuration settings

| Configuration Parameter | Value |
|---|---|
| Crossover Probability | 0.2 |
| Mutation Probability | 0.8 |
| Generations | 100 |
| Refactoring Range | 50 |

| Configuration Parameter | Value |
|---|---|
| Population Size | 50 |

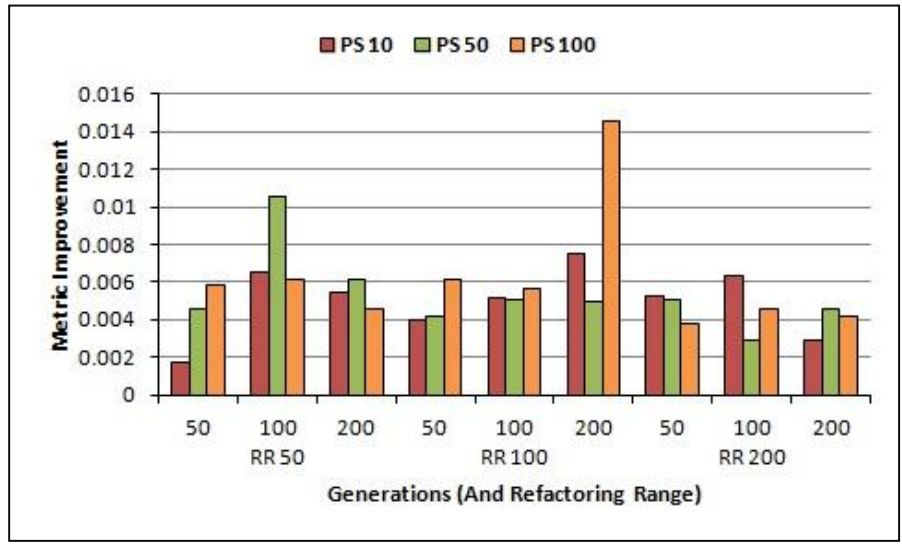

**Fig. 3.** Metric improvements for different configuration parameters

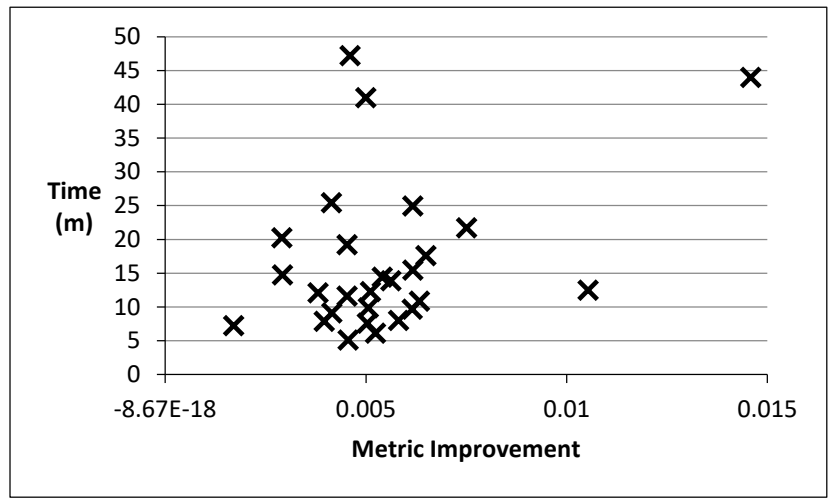

**Fig. 4.** Improvements mapped against time taken for different configuration parameters

In the first experiment, each metric is run as an individual fitness function with a genetic algorithm using the configuration parameters outlined in Table 4. The metrics are run with each of the input programs five times, giving an overall average improvement value. The average values are then compared for each metric to find the most volatile metrics with the available refactorings in the tool. In the second experiment, a set of metric functions are constructed using the results from the first, by excluding the metrics that have the least effect. The relevant metrics are split into three functions in order to be used as separate objectives in a multi-objective genetic algorithm. To compare the multi-objective approach with a mono-objective analogue, the three objectives are used as separate metric functions in different runs of the mono-objective algorithm. Each objective with the mono-objective search is run six times for each of the five inputs, giving 30 runs of the search. Likewise, the multi-objective genetic algorithm with the three objectives is run six times for each input. Therefore, across all four different search approaches, there are 120 tasks run.

For each objective, the mono-objective genetic algorithm is run using the configuration parameters from Table 4 for each input, and the average metric improvement is calculated for the top solution across the different inputs. For the purposes of this study, we are not interested in whether the multi-objective approach can generate a single solution with comparable results across all three objectives, but in whether each separate objective can be comparable. Therefore, the best solutions in the final population for each individual objective are found and the average improvements are calculated across the different inputs. In order to aid in finding the top scores for each objective in the final population of the multi-objective tasks, the search has been modified in this experiment to update the relevant results files to state that they contain the highest score for the corresponding objective, circumventing the need to manually check the scores in each solution.

The metric changes are calculated using a normalization function. The function finds the amount that a particular metric has changed in relation to its initial value at the beginning of the task. These values can then be accumulated depending on the direction of improvement of the metric and the weights given to provide an overall value for the metric function or objective. A negative change in the metric will be reflected by a decrease in the overall function/objective value. In the case that an increase in the metric denotes a negative change, the overall value will still decrease, ensuring that a larger value represents a better metric value regardless of the direction of improvement. For the experiments used in this paper, no weighting is applied to any of the metrics used. The directions of improvement used for each metric is defined in Table 5, where a plus indicates a metric that will improve with an increase and a minus indicates a metric that will improve with a decrease. Equation 1 defines the normalization function used, where $C_m$ is the current metric value and $I_m$ is the initial metric value. $W_m$ is the applied weighting for the metric and $D$ is a binary constant that represents the direction of improvement of the metric. $n$ represents the number of metrics used in the function.

$$\sum_{m=o}^{n} D \cdot W_m \left(\frac{C_m}{I_m} - 1\right) \quad (1)$$

## 4 Results

Table 5. Average metric gains

| Metrics | Direction | Average Metric Gain |
|---|---|---|
| Class Design Size | + | 0 |
| Number Of Hierarchies | + | 0 |
| Number Of Files | + | 0 |
| Average Number Of Ancestors | + | 0.0009662 |
| Number Of Children | + | 0.0009662 |
| Aggregation | + | 0.0028846 |
| Functional Abstraction | + | 0.00878788 |
| Number Of Polymorphic Methods | + | 0.00640564 |
| Abstractness | + | 0.0034176 |
| Inner Class Ratio | + | 0.0028846 |
| Lines Of Code | - | 0.0034388 |
| Data Access Metric | + | 0.07267708 |
| Direct Class Coupling | - | 0.011253 |
| Cohesion Among Methods | + | 0.0335982 |
| Number Of Methods | - | 0.047224824 |
| Weighted Methods Per Class | - | 0.07551 |
| Abstract Ratio | + | 0.06006748 |
| Referenced Methods Ratio | + | 0.02487444 |
| Visibility Ratio | - | 0.02984252 |
| Class Interface Size | + | 0.10246376 |
| Static Ratio | - | 0.17167356 |
| Final Ratio | + | 0.60217196 |
| Constant Ratio | + | 0.24485396 |

Table 5 gives the average quality gains conceived by each individual metric across all of the inputs. They are grouped into metrics that have a similar level of volatility. Three of the metrics, Class Design Size, Number Of Hierarchies and Number Of Files, showed no improvement at all. These metrics are more abstract, relating to the project design and class measurements as opposed to other metrics meas-uring more low level attributes like methods and fields. The most volatile metrics captured in the bottom group all relate to more low level aspects of the code. The metric functions used in experiment two were taken from the metric groups derived in Table 5. The least volatile metrics from the top two groups were left out and the remaining metrics were split into three individual objectives to be used in a multi-objective setup by using the three remaining groupings of metrics to each represent an objective. These particular groupings are informed by the average quality gains, with similarly volatile metrics being grouped together, although these groupings are used more as example objectives for the current experiment. These three groups of metrics may be combined to represent an overall improvement function for a generalized measure of software quality, with the average quality gain values across numerous different input programs informing its composition. Table 6 gives the list of metrics associated with each objective.

**Table 6.** Individual objectives derived from metric experimentation

| Objective 1 | Objective 2 | Objective 3 |
|---|---|---|
| Class Interface Size | Data Access Metric | Aggregation |
| Static Ratio | Direct Class Coupling | Functional Abstraction |
| Final Ratio | Cohesion Among Methods | Number Of Polymorphic Methods |
| Constant Ratio | Number Of Methods | Abstractness |
|  | Weighted Methods Per Class | Inner Class Ratio |
|  | Abstract Ratio | Lines Of Code |
|  | Referenced Methods Ratio |  |
|  | Visibility Ratio |  |

Fig. 5 and Table 7 compare the average objective values with the separate mono-objective runs against the values generated with the multi-objective approach. The values for objective one were the most disparate with the largest ranges of results. The mono-objective approach for objective 1 and objective 2 yielded improvements 1.2 and 1.3 times greater than the multi-objective approach, respectively. The other objective was slightly better with the multi-objective approach, though both improvement values where relatively small. The objective values for the two search approaches with the first and second objective were compared using a two-tailed Wilcoxon rank-sum test (for unpaired data sets) with a 95% confidence level ($α = 5\%$). The multi-objective values were found to not be significantly lower than the mono-objective values in either case.

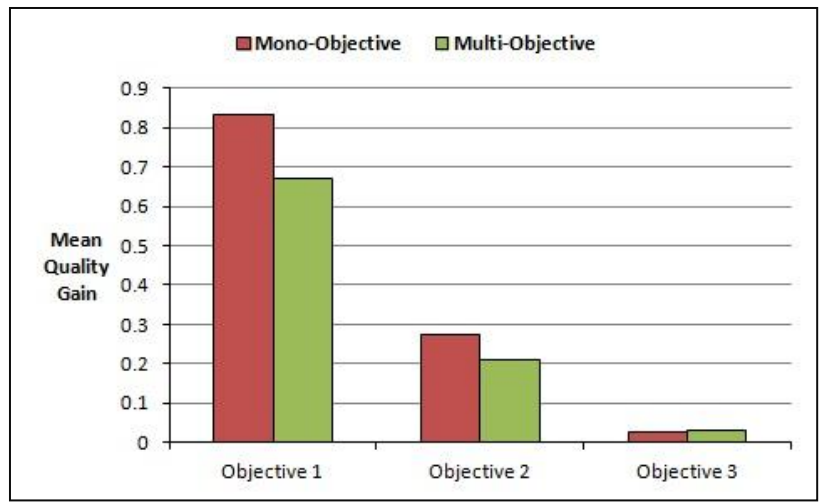

**Fig. 5.** Mean metric gains for each objective in a mono-objective and multi-objective setup

**Table 7.** Individual objective metric gains for mono-objective and multi-objective optimization

|  | Objective 1 | Objective 2 | Objective 3 |
|---|---|---|---|
| Mono-Objective | 0.8335831 | 0.2732774 | 0.028064733 |
| Multi-Objective | 0.672707033 | 0.210753367 | 0.028501433 |

The execution times for the two approaches were also compared to analyze how much more time is needed in the multi-objective approach to handle the three objectives simultaneously. Fig. 6 and Fig. 6 compare the overall times taken for the mono-objective and multi-objective approaches. In Fig. 6, the overall times taken for each individual objective of the mono-objective search are compared with the overall time taken to run the three objectives in the multi-objective approach. Fig. 6 compares the overall time taken to run all three objectives in the mono-objective approach against the multi-objective counterpart. It stacks the times for each separate objective in the mono-objective search to show the influence of each one on the time. The average time taken for the mono-objective algorithm to run for each objective was 3 hours, 46 minutes and 17 seconds. For the multi-objective approach to run for all the inputs it took 3 hours, 14 minutes and 49 seconds, a reduction against the mono-objective average of 31 minutes and 28 seconds. For the mono-objective approach to run the inputs for all three objectives would have taken over 11 hours, meaning 71.3% of time is saved running one multi-objective search against running three separate mono-objective searches.

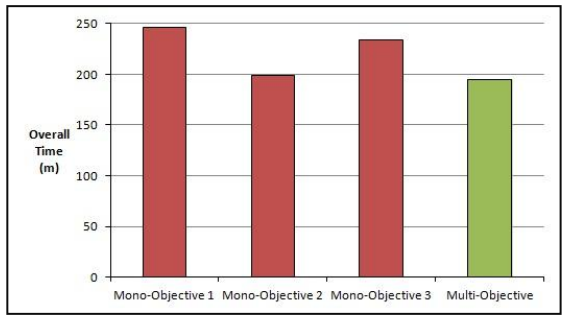

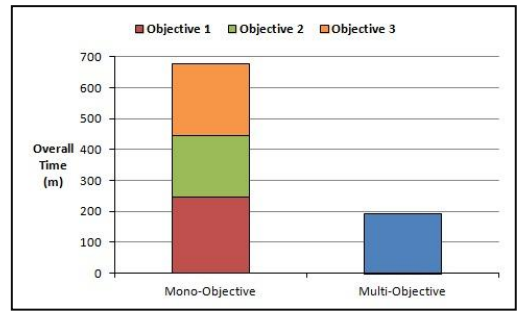

**Fig. 6.** Overall time taken to run each objective of the mono-objective approach and the multi-objective approach

**Fig. 7.** Overall time taken for each approach, with each objective of the mono-objective approach stacked on top of each other

## 5 Discussion

Of the metrics tested, three of the more abstract metrics showed no improvement. Although class level refactorings do exist in the MultiRefactor tool, they will be less likely to be applied due to the conditions necessary to apply them without modifying the program functionality. Likewise, the most volatile metrics all relate to more low level aspects of the code. It seems that these types of software metric may be more useful for driving change in an automated refactoring system due to the increased likelihood that structure level refactorings will be able to affect them.

To address **RQ2** and the answer the hypotheses constructed, statistical tests were used to decide whether the data sets were significantly different. While the other objective was better with the multi-objective approach, the statistical test was run for the first and second objectives where the multi-objective approach was worse. The values in the multi-objective approach were not significantly worse than in the mono-objective approach for either objective, thus rejecting the null hypothesis $H1_0$. In none of the three cases did the multi-objective approach take longer to run than the mono-objective approach, thus rejecting the null hypothesis $H2_0$. The experiments conducted suggest that this fully automated approach may be feasible and can allow for multiple separate objectives to be considered in a single run within an acceptable amount of time, although the improvement of a subset of these objectives may take a hit.

## 6 Related Work

The term SBSE was first coined by Harman and Jones in 2001 [1]. Further research in the area was identified, as well as open problems in 2007 [29]. Clarke et al. [30] discussed ways to apply metaheuristic search techniques to software engineering problems and proposed other aspects of software engineering to apply them to in 2003. There are literature reviews on the subject [31, 32]. Numerous tools have been proposed that can automate the maintenance process of software refactoring to some extent, although many are limited, and not all are fully automated. Many of the proposed tools isolate design smells in the code using detection rules [3–8]. Most of the tools using this approach have focused on a limited amount of detection rules to isolate certain types of design smell, due to the uncertainty involved in constructing these metric based detection rules. Other tools use metrics to determine ideal refactorings to make to the code that will improve the quality and remove design smells as a by-product of the process [9–11].

More recent research has explored the use of multi-objective techniques. White et al. [12] used a multi-objective approach to attempt to find a tradeoff between the functionality of a pseudorandom number generator and the power consumption necessary to use it. De Souza et al. [13] investigated the human competitiveness of SBSE techniques in four areas of software engineering, and used mono-objective and multi-objective genetic algorithms in the study. Ouni et al. [14] created an approach to measure semantics preservation in a software program when searching for refactoring options to improve the structure, by using the NSGA-II search. Ouni et al. [15] then explored the potential of using development refactoring history to aid in refactoring a software project by using NSGA-II. Ouni et al. [17] also expanded upon the code smells correction approach of Kessentini et al. [16] by replacing the genetic algorithm used with NSGA-II. Mkaouer et al. [18] experimented with combining quality measurement with robustness using NSGA-II to create solutions that could withstand volatile software environments. Mkaouer et al. [19, 20] also used the successor algorithm to NSGA-II, NSGA-III, to experiment with automated maintenance. These studies only suggest refactoring sequences to be applied, and do not check the applicability of the refactorings.

## 7 Conclusion

In this paper we have presented the MultiRefactor approach and associated automated refactoring tool containing both mono-objective and multi-objective search techniques. Six separate search techniques are available as well as 23 different metrics and 26 refactorings. The tool works with Java source code (as well as accompanying library files) as input and is a fully automated tool that can generate refactored, compilable Java code as an output, along with information about the refactoring process. The tool is highly configurable, allowing the user to set up different tasks with different sets of metrics to use and different refactorings to activate. The available search techniques have numerous configurable properties to be set, influencing how the search process will work. No other known refactoring tool currently allows the user to use multi-objective techniques to improve the software without having to manually apply the refactorings.

Two experiments were run to test various aspects of the approach. The configuration parameters of the genetic algorithm were tested to analyze the effect that they can have on the refactoring process and to deduce what settings can have a better tradeoff between metric improvement and time taken. Each of the available metrics were then tested with the genetic algorithm across a number of real world, open source Java programs to find the least volatile metrics interacting with the available refactorings. It was found that the more low level metrics produced greater average improvements compared to the more abstract, class level metrics. The results of this experiment were then used to construct metric functions to compare a mono-objective refactoring approach against a multi-objective approach. The more volatile metrics were split into three separate objectives to see if the multi-objective approach could generate comparable results to the mono-objective counterparts. The individual mono-objective approaches gave better results for two out of the three objectives but the multi-objective approach managed to generate suitable improvements for all of the objectives and took less time than each mono-objective approach, with the single multi-objective run taking 71% less time than the three combined mono-objective runs.

**Acknowledgments.** The research for this paper contributes to a PhD project funded by the EPSRC grant EP/M506400/1.